\documentclass[twocolumn,showpacs,preprintnumbers,amsmath,amssymb]{revtex4}
\usepackage{graphicx}
\usepackage{dcolumn}
\usepackage{bm}

\begin{document}

\title{Electronic Structure and Magnetic Properties of Y$_{4}$Co$_{3}$ }

\author{T. Jeong}
  
\affiliation{
Department of Physics, University of California, Davis, California 95616
}

%\date{\today}

\begin{abstract}
The electronic structure of Y$_{4}$Co$_{3}$ has been
studied based on the density functional theory within the
local-density approximation.
The calculation indicates that Y$_{4}$Co$_{3}$ is very
close to ferromagnetic instability.
The Fermi surfaces are composed mainly
of $3d$ electrons of Co and $4d$ electrons of Y.

\end{abstract}


\maketitle

\section{\label{sec:level1}Introduction}

The intermetallic Y$_{4}$Co$_{3}$ has attracted much attention 
since the coexistence of magnetism and superconductivity were 
discovered in this compound
and it is the first itinerant band magnetic superconductor and reported 
to be weakly ferromagnetic \cite{kol,berthet, gratz}.
Y$_{4}$Co$_{3}$ has an unusual interplay of 
magnetism and superconductivity.
The compound Y$_{4}$Co$_{3}$ is very interesting because neither Co nor Y as
metals are superconductors.
Magnetism (T$_{c} \approx 5 $K) precedes superconductivity 
(T$_{sc} \approx 2.5 $K) and the inverse of the behavior has shown most often 
elsewhere \cite{lewicki, sarkissian}.
It is also interesting that 
 in the Y-Co system the magnetic moment per Co atom tends 
to zero for YCo$_{2}$, which is a Pauli paramagnet, and the 
compositions close to Y$_{4}$Co$_{3}$ shows a very small magnetic moment 
about $0.012 \mu_{B}$ per Co atom which is localized 
on a specific cobalt site \cite{gratz, yamaguchi}.

There is experimental evidence from NMR that $d$-type cobalts 
enclosed in triangular prisms of Y atoms are responsible 
for the magnetic properties\cite{figiel}. 
And the small Co-Co distance prevents magnetism in phases 
neighbouring to Y$_{4}$Co$_{3}$, a rather different structure 
of Y$_{36}$Co$_{28}$(Y$_{9}$Co$_{7}$) phase, 
with the isolated Co atoms along the c-axis, is responsible 
for the magnetic properties and strong development of magnetic 
correlations \cite{sarkissian}.
The specific heat experiments showed that the phonon
contribution is mainly due to Y lattice but the 
electronic contribution is due to the Co density of electron states
\cite{lewicki}.
Some important features of this type magnetic superconductor remain 
to be explained theoretically and 
the preliminary electronic structure study of Y$_{4}$Co$_{3}$ is 
essential to understand the fascinating properties of this compound.
In table \ref{tableweak} we summarized the physical characteristics 
of Y$_{4}$Co$_{3}$ and other weak ferromagnets. 
From the data we can see that
Y$_{4}$Co$_{3}$ has indeed weak itinerant ferromagnetic properties.

%\begin{figure}
%\includegraphics[height=7.5cm,width=8.5cm,angle=0]{fig1a.eps}
%\vskip -10mm
%\includegraphics[height=7.5cm,width=8.5cm,angle=0]{fig1b.eps}
%\caption{The crystal structure of Y$_{4}$Co$_{3}$.
% The larger atoms are Y, and
% the smaller one are Co.} 
%Upper panel is a top view.}
%\label{structure} 
%\end{figure} 
 
Until now there is no detailed electronic band structure studies for   
Y$_{4}$Co$_{3}$, therefore 
as a first step  the electronic 
structure investigaion based on the density functional theory is important 
to understand the nature of this peculiar compound.
In this work,
the precise self-consistent full potential  
local orbital minimum basis band structure scheme (FPLO) are 
employed to investigate thoroughly the electronic and magnetic 
properties of Y$_{4}$Co$_{3}$ based on the density functional theory.
We focus on the studies of the effect of magnetism on the
 band structure and Fermi
surfaces and compare with experiment results.

\section{Crystal Structure}

Y$_{4}$Co$_{3}$ is in the hexagonal Co$_{3}$Ho$_{4}$ 
crystal structure ( space group 
$P6_{3}/m$, $\#$176)
% which is shown in the Fig. \ref{structure}. 
This crystal structure is descibed in detail by Yvon 
Yvon {\it et al.}\cite{yvon}.
An unusual feature of this crystal structure is the 
existence of three inequivalent crystallographic 
positions of cobalt.
Y$_{4}$Co$_{3}$ can be viewed as quasi-infinite triangular columns 
of trigonal Y prisms running along a $\bar{6}$ axis which are made up 
of two types of Y atom(Y(1), Y(2)), both
 at site $6(h)$(the site symmetry is m), 
and are centered by two
types of Co atom at site $6(h)$ and $2(d)$ (the site symmetry is m and 
$\bar{6}$ respectively). 
In Co$_{3}$Ho$_{4}$ structure a third atom site is 
located on the ${\bf c}$-axis, i.e., in the channels formed 
by six groups of triangular Ho prism columns(at the site  $2(b)$, 
site symmetry $\bar{3}$ ).
Y$_{4}$Co$_{3}$ crystal structure has inversion symmetry but
nonsymmorphic.
For our calculation we used 
the experimental lattice constants, $a=11.527 \AA$  and
$c=4.052  \AA$.
There are three formula units per cell.

\begin{table}[b]
\caption{\label{tableweak}Y$_{4}$Co$_{3}$ and other weak ferromagnets  }
\vskip 5mm
\begin{center}
\begin{tabular}{|l|c|c|c|c|}
\hline
physical properties  &  Y$_{4}$Co${3}$   &   Sc$_{3}$In & ZrZn$_{2}$ \\
\hline
$\chi_{0}$(emu/g)   & 2.25 $\times 10^{-6}$  & 0.4 $\times 10^{-6}$ 
  & 2.16 $\times 10^{-6}$  \\
T$_{c}$ (K)   &5.0     & 21.3   & 5.5 \\
moment ($\mu_{B}$/atom) & 0.012/Co    & 0.045/Sc  & 0.12/Zr  \\
$\gamma$ (mJ/mol K$^{-2}$)  &  3.4   & 33.6 & 47.0 \\

\hline
\end{tabular}
\end{center}
\end{table}

\begin{table}[b]
\caption{\label{table2} distances between atoms of Y$_{4}$Co$_{3}$ }
\vskip 5mm
\begin{center}
\begin{tabular}{|l|c|c|}
\hline
Atom    &   Atom   &   distance $(\AA)$   \\
\hline
Co(h)   &   Co(h)  &  4.05   \\
Co(b)   &   Co(h)  &  4.61   \\
Co(d)   &   Co(h)  &  2.34   \\
Co(b)   &   Y(1)  &  2.96   \\
Co(d)   &   Y(1)  &  2.91   \\
Co(h)   &   Y(1)  &  2.83  \\
Y(1)    &   Y(2)  &  3.32   \\

\hline
\end{tabular}
\end{center}
\end{table}

\section{Method of Calculations}

We have applied the full-potential 
nonorthogonal local-orbital minimum-basis (FPLO) scheme within the local 
density approximation (LDA).\cite{koepernik}
In these scalar relativistic calculations we 
used the exchange and correlation potential of Perdew and Wang.\cite{perdew}
Y $4s, 4p, 4d, 5s, 5p$, and Co $3s, 3p, 3d, 4s, 4p$ states 
 were included as 
valence states. All lower states were treated as core states.
We included the relatively extended semicore 3s, 3p states of Co and
3s, 3p, 3d states of Y as band states
because of the considerable overlap of these
states on nearest neighbors.
This overlap would be otherwise neglected in our FPLO scheme.
 The spatial extension
 of the 
basis orbitals, controlled by a confining potential $(r/r_{0})^4$, was 
optimized to minimize the total energy. The self-consistent potentials were 
carried out on a k mesh of 16 k points in each direction of the Brillouin zone,
which 
corresponds to 132 k points in the irreducible zone.

\section{Results}
\subsection{Band Structure and Density of States}

\begin{figure}
\includegraphics[height=8.5cm,width=8.5cm,angle=-90]{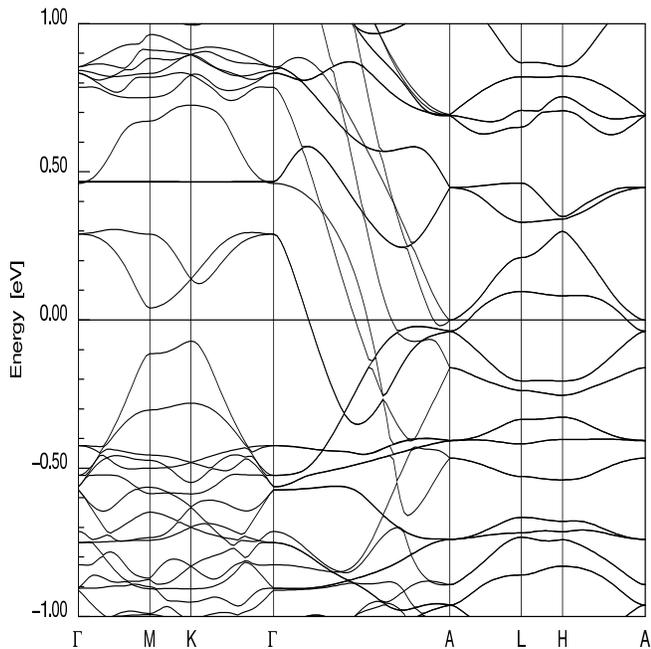}
\caption{The LDA bandstructures of 
Y$_{4}$Co$_{3}$  near Fermi level.} 
\label{pmband}
\end{figure}

\begin{figure}
\includegraphics[height=8.5cm,width=8.5cm,angle=-90]{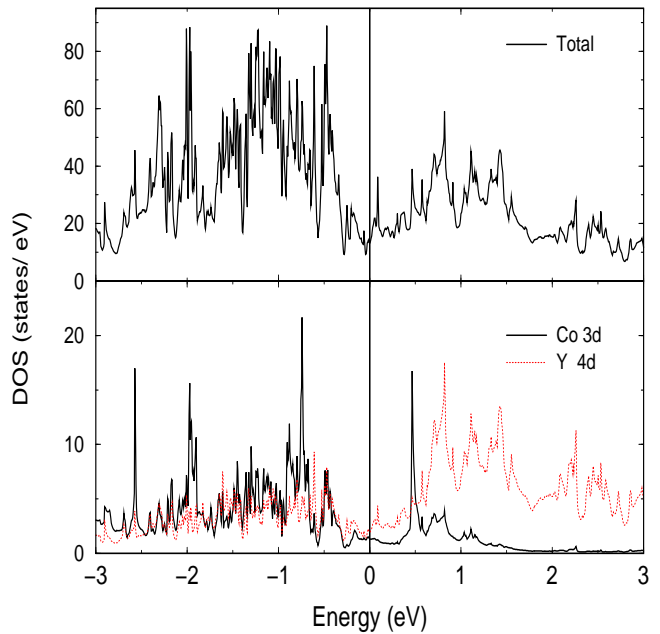}
\caption{The total and atom-projected density of states of Y$_{4}$Co$_{3}$  . 
The contribution of Co 3d and Y 4d to density of states.
} 
\label{dos}
\end{figure} 

\begin{figure}
\vskip 5mm
\includegraphics[height=6cm,angle=-0]{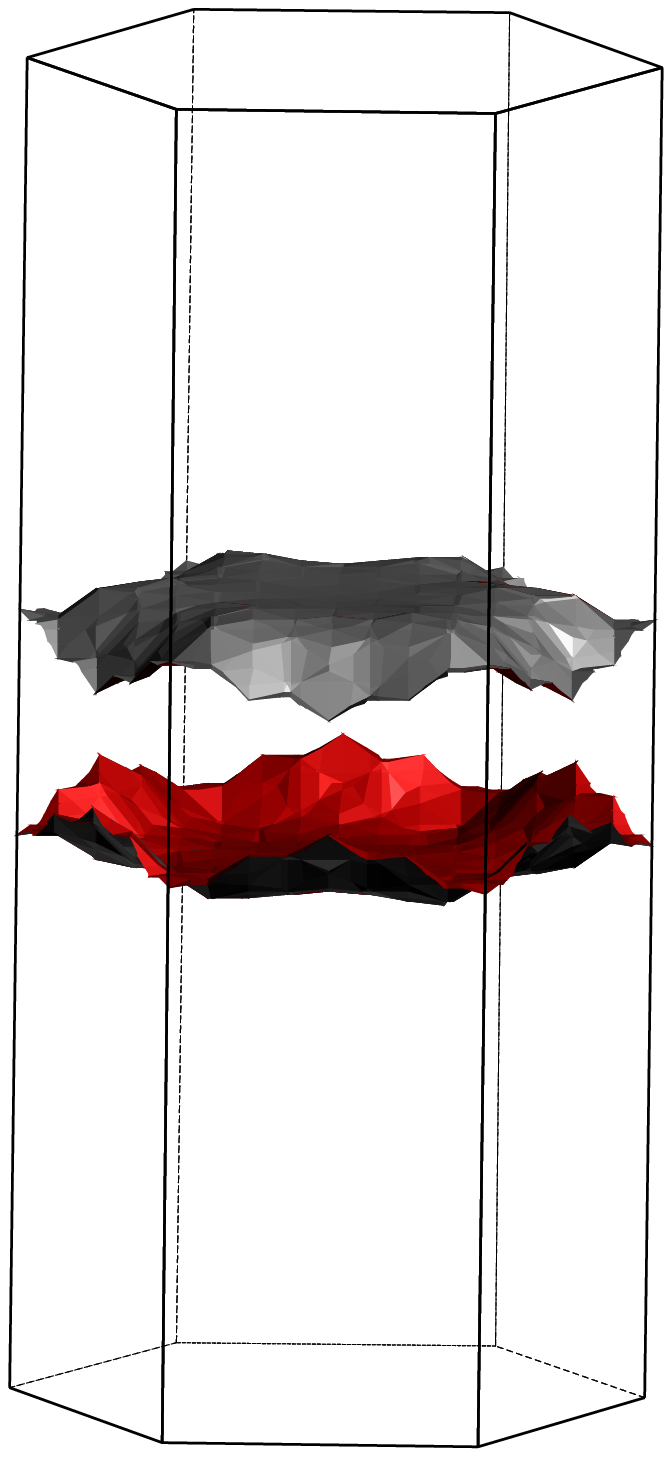}
\includegraphics[height=6cm,angle=-0]{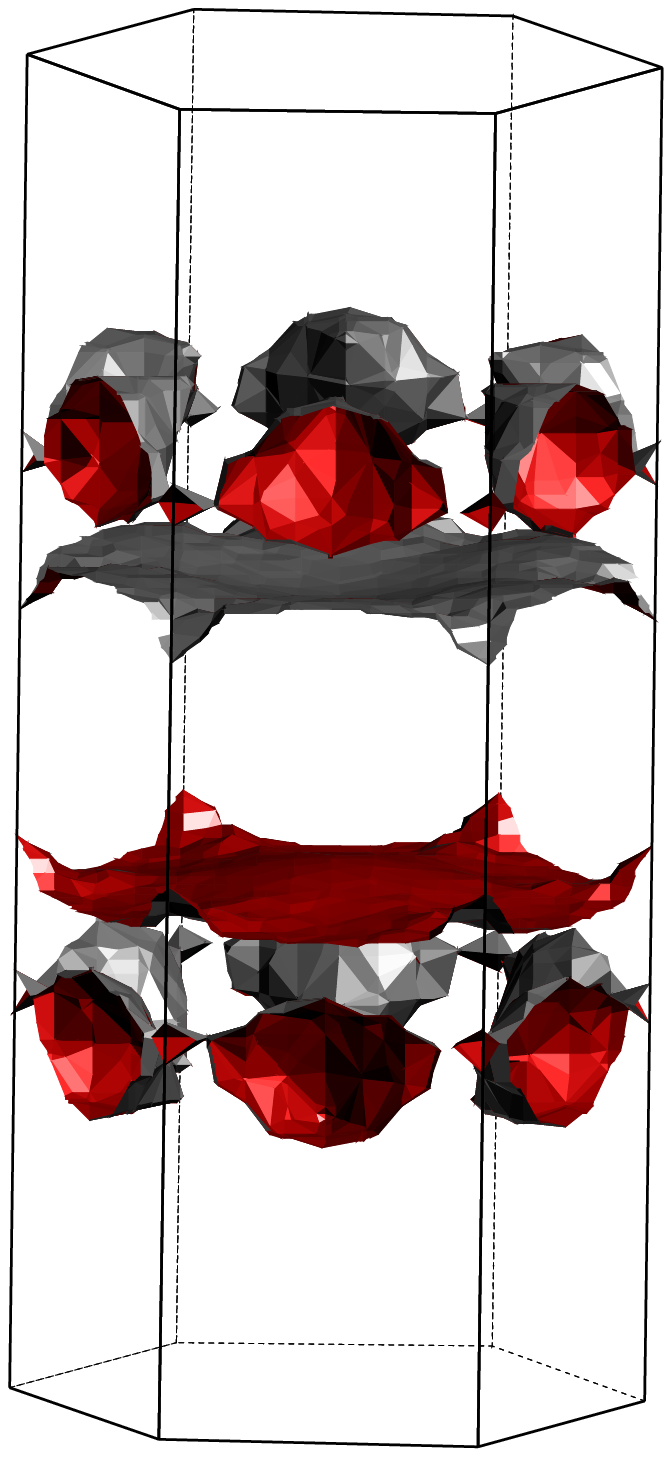}
\vskip 5mm
\includegraphics[height=6cm,angle=-0]{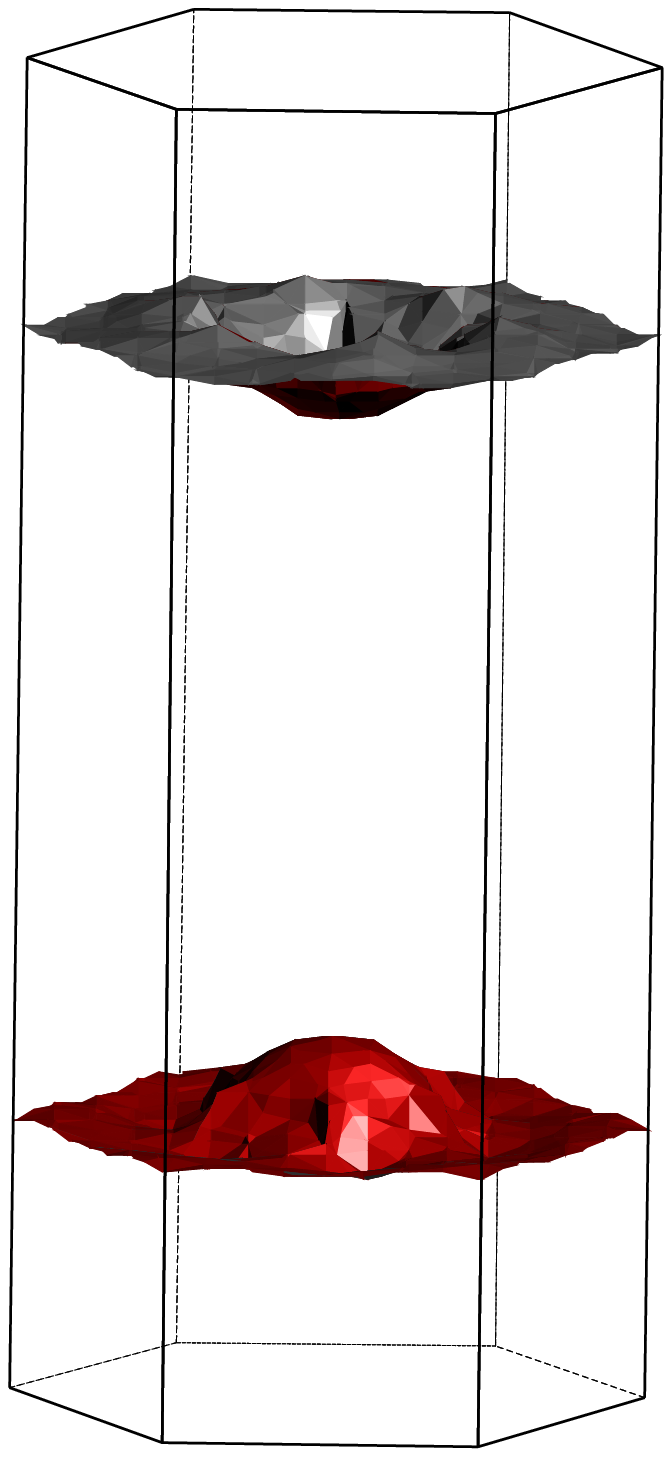}
\includegraphics[height=6cm,angle=-0]{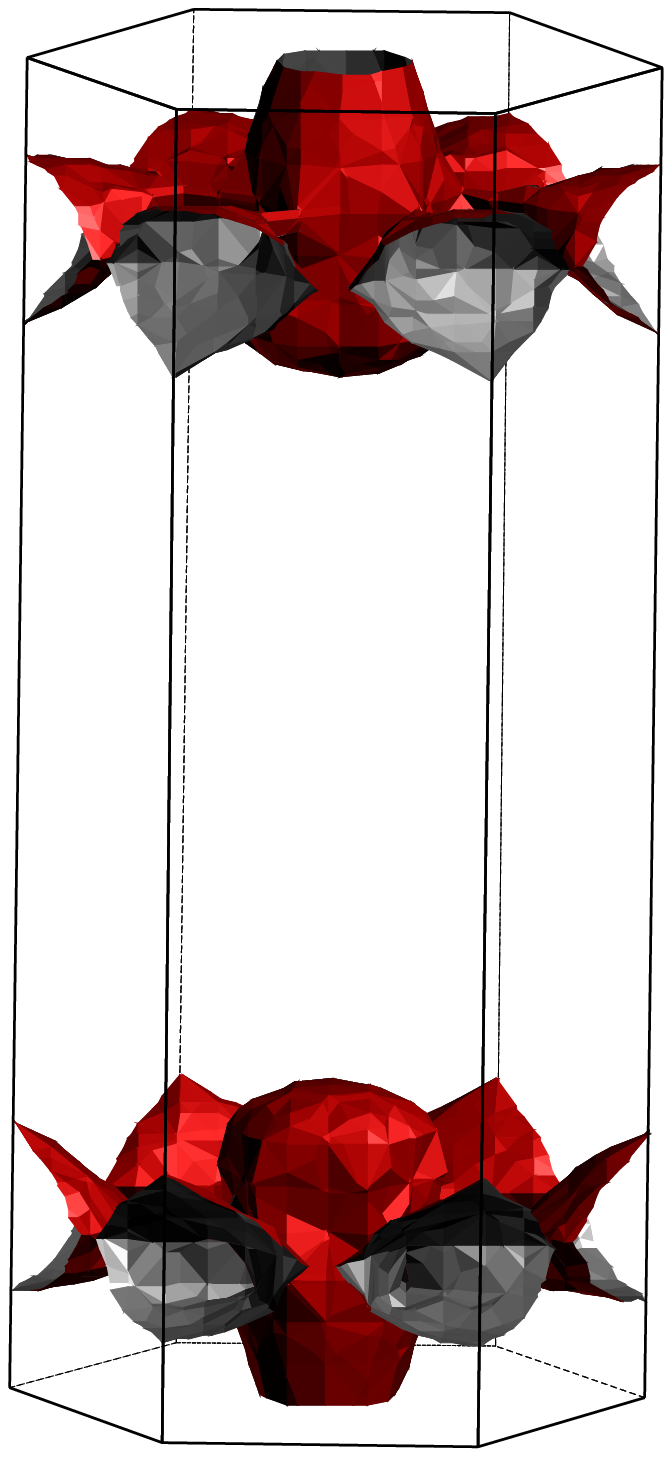}
\vskip 5mm
\includegraphics[height=6cm,angle=-0]{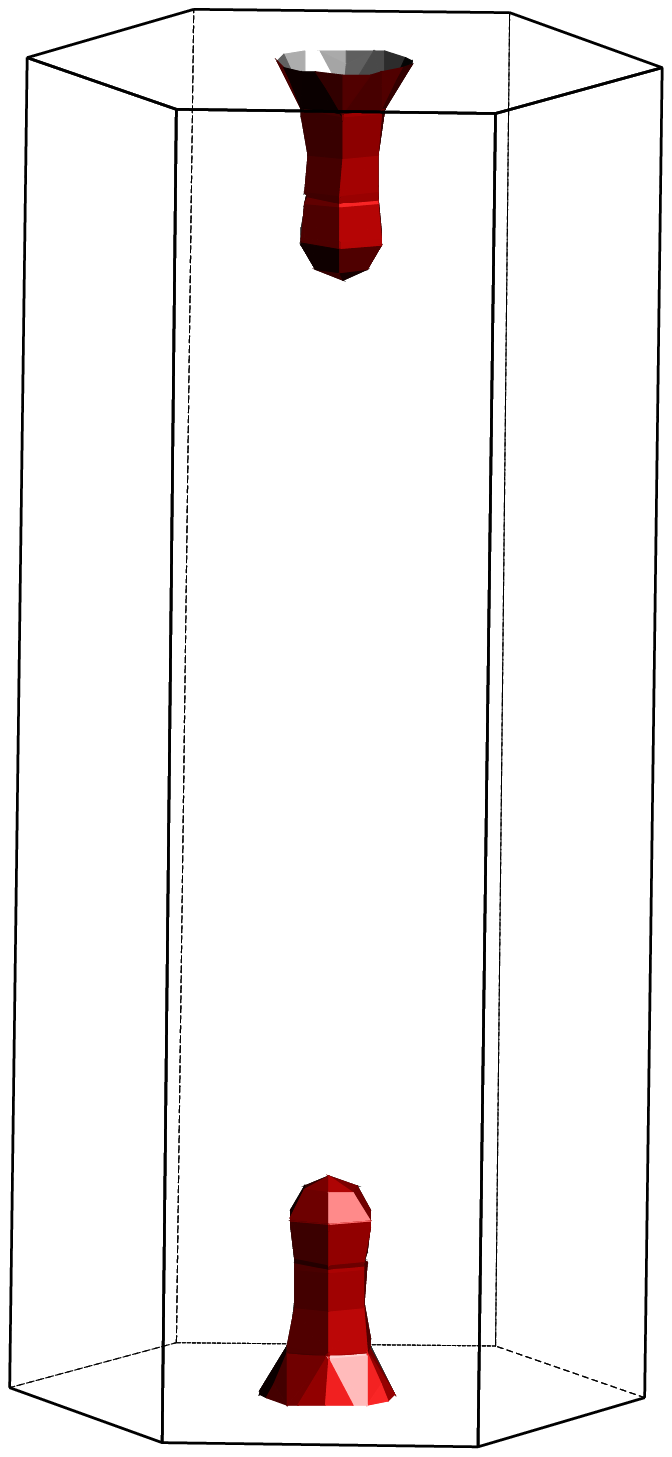}
\caption{(Color online)
 Fermi surfaces of Y$_{4}$Co$_{3}$ (from top, bands 257, 258, 259, 260, 261 ).
The $\Gamma$ point corresponds to the center of the hexagonal prism; 
the $A$ point is the midpoint of the lower and upper hexagon. 
 }
\label{FS}
\end{figure}

The electronic and magnetic studies of Y$_{4}$Co$_{3}$
 based on density functional theory 
help to characterize this interesting compound.  
The Fermi energy lies in a band of $d$ electrons  
of the Y atoms and more suppressed by moments
on those sites than by the magnetic character of some of the 
Co atoms. 
The Y 3$d$ bands are almost filled and this leads to both 
a small moment and small spin splitting.

Around the Fermi energy Y 4d and Co 3d states are mixed,
so the magnetism and superconductivity are strongly 
linked due to the hybrized Y-Co bands. 
The most intriguing observation in the band structure is the 
0.15 eV energy gap along $\Gamma-M-K-\Gamma$ symmetry line which is 
between the bands of Y 4d states.
Also there is very flat band located 0.43 eV above Fermi level along 
these lines which are the Co 3d character. 
There is a flat band along 
$L-H$ line at 0.1 eV above Fermi level which is the Y 4d character.
Density of states (DOS) is shown in Fig. \ref{dos}.
The DOS near the Fermi level arises from Co $3d$ bands
hybridized with Y $4d$ states.
Lewicki  {\it et al.}\cite{lewicki} measured the linear specific heat
coefficient for Y$_4$Co$_3$ 
of $\gamma$=3.4 mJ/K$^{2}$ mole(formular unit). 
The calculated value of $N(E_{F})$= 4.11 states/eV for Y$_4$Co$_3$
corresponds to
a bare value $\gamma_b$= 9.62 mJ/K$^{2}$ mole(formular unit), which 
is larger than the experimental value.  

\subsection{Ferromagnetic Instability and Fermi Surfaces}

Density functional calculations are usually reliable in calculating 
the instability to ferromagnetism.
The enhanced susceptibility\cite{janak} is given by 

\begin{eqnarray}
\chi =\frac{\chi_{0}}{[1-N(E_{F})I]} \equiv S\chi_{0}.
\label{susceptibility}
\end{eqnarray}

where $\chi_{0}=2\mu_{B}^{2}N(E_{F})$ is the bare
susceptibility obtained directly from the  band
structure and $I$ is the Stoner exchange interaction constant.
 The calculation of $I$ 
resulted from fixed spin moment calculations\cite{mohn},
 in which 
the energy $E(m)$ is calculated subject to the moment 
being constrained to be $m$. The behavior at small m 
is $E(m)=(1/2)\chi^{-1}m^{2} $ from which $I=0.18$ eV can be 
extracted form Eq. \ref{susceptibility}.
This gives $IN(E_{F})=2.2 $, larger than unity, corresponding 
to a ferromagnetic instability.

The corresponding Fermi surfaces (FS) of Y$_{4}$Co$_{3}$
is shown in Fig. \ref{FS}.
The $\Gamma$ point is located at the center of the hexagonal prism 
and the $A$ point is the midpoint of the lower and upper hexagon. 
Both the 4$d$ electrons of Y and the 3$d$ electrons of Co 
contribute to the Fermi surface. 
Due to the band gap along the symmetry lines $\Gamma MK$, there 
is no FS at the center of the hexagonal prism.
The crossing bands near the symmetry line A make some complicated 
Fermi surfaces.
In the bottom panel, 
we can observe the cylindrical tube-like Fermi surfaces.
The non-magnetic component determines the position of the 
Fermi level at the bottom or top edge of the $d$ band. The fact 
that $\chi_{0}$ for Y$_{4}$Co$_{3}$ is so close to the corresponding 
value for Y shows that it is the $5s$ electrons of Y 
which are transferred mainly to $3d$ orbitals of Co, and thus 
filling them while leaving the $4d$ electrons of Y almost inact. 
Therefore the Fermi surface should be composed mainly
of $3d$ electrons of Co and $4d$ electrons of Y.

\section{Discussion and Conclusions}

The weak ferromagnetic compounds exhibit the following typical 
physical properties;
(1) a low saturation moment per transition metal(TM) atom at 0K
($<0.4 \mu_{B}$/TM atom). 
(2) a low magnetic order-disorder phase transition temperature 
($T_{c} <$ 45K).
(3) a large high-field magnetic susceptibility at 0K.
(4) a large coefficient of the term linear in temperature in the specific 
heat at low temperature.
(5) $T^{5/3}$ power law behavior of the resistivity over a wide 
temperature range around $T_{c}$. 
(6) Curie-Weiss behavior of the magnetic susceptibility in the
paramagnetic state\cite{spin}.
From the experimental reports we can 
conclude that Y$_{4}$Co$_{3}$ joins the weak ferromagnets group 
such as ZrZn$_{2}$, Sc$_{3}$In, and Ni$_{3}$Al.
According to the the band magnetism, 
to get a ferromagnet with small magnetic moment 
in a narrow 3d band 
it must be either almost filled or almost empty. 
The examples of the former is Ni$_{3}$Al, Y$_{4}$Co$_{3}$ and 
the examples of the latter is ZrZn$_{2}$, Sc$_{3}$In. A charge transfer 
from a non magnetic to a magnetic element with a partly 
filled $d$ shell should take place, because 
the stable phases are antiferromagnetic 
and ferromagnetic with relatively large moments per site 
when the shell is almost half filled .
Such an 
observation could explain why weak ferromagnetism occurs mainly 
in intermetallic compounds, such as ZrZn$_{2}$,Sc$_{3}$In and Ni$_{3}$Al. 

Kolodziejczyk {\it et al.}\cite{kol} measured the resistivity of 
Y$_{4}$Co$_{3}$, in which  
the resistivity $\rho(T)$ of Y$_{4}$Co$_{3}$ follows that 
$\rho(T)= T^{2}$ in the lower temperature (5-17K)and 
$\rho(T)= T^{5/3}$ in the higher temperature(20-70 K).
This results agrees very well with the theory of Ueda and Moriya\cite{ueda}.
We can expect the T$^{2}$ law to persist also well below T$_{c}$, 
but such a behavior of the resistivity is suppressed by superconductivity.
We would expect that the band undergoing spin 
fluctuation should be the hybrised $3d-4d$ band. This is because 
the distance between neighboring Co atoms belonging to 
different positions are large while the Y-Co distances are small.
The distance Y(1)-Y(2) is approximately 3.32 $\AA$ and the distance 
between atoms in the cluster forming a prism is Co(d)-Co(h)=2.34$\AA$, 
which is smaller than the distance Co-Co along the 
a axis in hexagonal close pack structure of Co. The other cluster 
would be composed of Co atoms which occupy only some of the 
available positions and thus can form pairs or short chains along the 
c axis. The clusters of four Co atoms in the prism form 
molecular orbitals which mix with the Y $4d$ and $5s$ states. 
The low spin state could be the lowest energy state for this molecule. 
Additional reduction of the moment comes from the charge 
transfer from Y to Co, therefore the $3d$ band is almost filled 
and this leads to both a small magnetic moment and small spin splitting. 
 
%\section{Acknowledgments}

%The author acknowledge illuminating discussions with H.O.Lee.

\end{document}